\documentclass[preprint,aps,prb]{revtex4}
\usepackage{epsfig}
\usepackage{float}

\begin{document}
\title{
Aqueous solvation of methane from first principles}
\author{Lorenzo Rossato, Francesco Rossetto, and Pier Luigi Silvestrelli}
\address{Dipartimento di Fisica ``G. Galilei'',
Universit\`a di Padova, via Marzolo 8, I-35131 Padova, Italy
and DEMOCRITOS National Simulation Center, Trieste, Italy}

\date{\today}

\begin{abstract}
Structural, dynamical, bonding, and electronic
properties of water molecules around a soluted methane molecule
are studied from first principles. 
The results are compatible with experiments and qualitatively 
support the conclusions of recent classical Molecular Dynamics simulations
concerning the controversial issue on 
the presence of ``immobilized'' water molecules around 
hydrophobic groups:
the hydrophobic solute slightly reduces (by a less than 2 factor)
the mobility of many surrounding water molecules rather than immobilizing
just the few ones which are closest to methane, similarly to what
obtained by previous first-principles simulations of soluted methanol. 
Moreover, 
the rotational slowing down is compatible with that one predicted 
on the basis of the excluded volume fraction, which leads to a slower Hydrogen 
bond-exchange rate.
The analysis of
simulations performed at different temperatures suggests that 
the target temperature of the soluted system must be carefully chosen,
in order to avoid  artificial slowing-down effects.
By generating maximally-localized Wannier functions, a detailed 
description of the polarization effects in both solute and solvent molecules
is obtained, which better characterizes the solvation process.  
\end{abstract}
\maketitle
\vfill
\eject

\narrowtext

\section{Introduction}
Understanding the properties of water around hydrophobic solutes,
which are mainly determined by the Hydrogen bond (Hb) network,
is crucial since the rotational and translational motion of water
molecules in solution control the rates of important chemical
and biochemical processes,\cite{REVIEW,Bakker} such as protein folding 
and membrane assembly.
Basically, the hydrophobic effect represents the tendency of
apolar groups to associate in aqueous solutions and
minimize the total surface that is exposed to water;
instead polar groups can participate in Hbs 
with water molecules. 

In particular, the aqueous solution properties of methane 
have long been of interest: in fact, methane is the simplest 
hydrocarbon molecule and represents a good model compound for understanding 
the properties of water molecules in the solvation shell of small apolar
solutes.
Methane-water systems play also an important role in various fields of
the natural sciences, ranging from fossil fuel chemistry to reactions
in the conversion of solar energy (thermochemistry), to methane
production from fermentation of farm wastes.\cite{Laaksonen}

It is now generally believed that the nature of hydrophobicity is size
dependent:\cite{Chandler} small apolar molecules such as methane have a
hydration free energy near ambient conditions that is largely entropic; 
that is, it depends more on the number of ways all of the water molecules
in the methane hydration shell can form Hbs rather than their energies.
Water molecules in the vicinity of small non-polar solutes 
are capable\cite{Chau} of rearranging themselves in such a way to
encapsulate the guest molecule and  
to regenerate any Hbs broken during the insertion of the solute, 
so that a non-polar solute can be
thought of as a "structure maker";\cite{Chau,Lambeth} it is then possible to 
describe hydrophobic hydration in terms of the directional, Hydrogen bonding
properties of water, which undergoes a restricted orientational 
freedom due to the inability of the solute to participate in Hbs, 
which makes certain orientations of the hydration shell water
molecules energetically unfavourable.\cite{Chau} 

In spite of the importance of the subject a detailed, microscopic
understanding of water dynamics in vicinity to an apolar solute
remains elusive and different mechanisms have been conjectured.
It was suggested\cite{Frank} that hydrophobic solutes enforce
the network of H-bonded water molecules around them and strongly decrease their
mobility, in such a way that they form rigid, icelike structures
(``icebergs'').
However, several experimental and computational studies showed that, at least
at low solute concentration around room temperature,
the structure and dynamics of the water molecules in the solvation
shell is similar to that of bulk 
water.\cite{SIMULATIONS,Laage09,Laage06,Sciortino,Qvist}
Nuclear Magnetic Resonance (NMR) and Dielectric Relaxation (DR)
experiments\cite{EXPT} indicated that the mobility of water
molecules in solutions containing hydrophobic solutes is decreased,
however, since these methods can only measure a response averaged
over all water molecules, 
it was not clear whether well-defined layers of ``immobilized''
water molecules exist or instead there are many molecules slightly
affected in their dynamical behavior.
Recent results by Rezus and Bakker,\cite{Bakker} obtained by using
femtosecond midinfrared spectroscopy (fs-IR) to investigate the 
hydration of hydrophobic groups on a subpicosecond time scale
(experiments\cite{Fecko} use "ultrashort"-infrared pulses 
with durations of the order of 50 fs),
have been interpreted as a confirmation that water
molecules surrounding hydrophobic groups
undergo a much slower orientational dynamics than the bulk liquid and
are therefore effectively immobilized; in particular, it is suggested that
each methyl
group is surrounded by 4 immobilized waters, supporting
the ``iceberg'' picture in a {\it dynamical} way: although the water molecules 
around hydrophobic groups have a liquidlike structures, their dynamics
would be icelike, which also explains why hydrophobic icebergs 
were not previously observed with structural methods.

The major experimental problem in the NMR measurements of water-methane
systems is due to the
low solubility of methane in water; in readily accessible proton spectra
a large proton water peak would effectively mask the signal from small 
amounts of dissolved methane.\cite{Laaksonen} 
Computer simulations can give very detailed information about the 
solvation structure, providing detailed mechanistic insights that 
cannot be directly obtained from experiments.
In one of the first classical Molecular Dynamics (MD) simulations 
of methane-water
systems, Laaksonen and Stilbs\cite{Laaksonen} found that the dynamics of the 
solvent water molecules is slowed down (by about a 1.5 factor in the
reorientational correlation time): in particular, the 
reorientational motion is sensitive to the presence of methane in
solution; the simulated translational diffusion of methanes as
solutes is also slowed down, but is higher than observed experimentally.
Moreover, methane acts nearly as a free rotor.\cite{Laaksonen}

Subsequent classical MD simulations\cite{Laage09,Laage06,Sciortino,Qvist} 
supported the idea that a moderate
slowing down of water dynamics around hydrophobic groups arises
from a steric effect, with a 
consequent decrease in the configurational space available
to water molecules around hydrophobic solutes.
Interestingly,
Stirnemann {\it et al.},\cite{Stirnemann} investigated the Hb 
dynamics of water in a series of
amphiphilic solute solutions (solute molecules possessing both hydrophobic 
and hydrophilic moieties) through classical MD simulations 
and analytic modeling:
they find that for most solutes the major effect in the hydration
dynamics comes from the hydrophilic groups that can retard the water
dynamics much more significantly (the retardation factor being about 1.4) 
than can hydrophobic groups by forming
strong Hbs with water; by contrast, hydrophobic groups are shown to have a
very moderate effect on water Hb breaking kinetics
via an excluded volume effect.\cite{Stirnemann}
This interpretation is in contrast with, for instance, that of Bakulin
{\it et al.}\cite{Bakulin} who studied solvation of tetramethylurea and
other amphiphilic solutes, and
attribute the retarded spectral relaxation
observed in the experimental spectra to a dramatic slowdown in the water
dynamics induced by the hydrophobic groups, due to the suppression
of water jumps between Hb acceptors: in particular, due to a steric effect,
the fifth water molecule
which is required to form a defect state in the tetrahedral surroundings
(with the temporary formation of a bifurcated Hb), 
cannot approach the H-bonded pair to initiate the molecular jump; as a
results, the rate of the jumping events sharply decreases, which, in
turn, strongly slows the rotation of the water molecules.\cite{Bakulin}
Therefore, the effect of hydrophobic groups would be twofold: they do 
not only exclude part of the volume available for switching but 
fundamentally change the Hb dynamics of the solvating water molecules; this
can be associated with the formation of a more "rigid" water-solute 
structure where the translational
mobility decreases; this explanation is supported by the suppressed
translational mobility of water and its excellent correlation to
Hb dynamics.\cite{Bakulin}
Bakulin {\it et al.}\cite{Bakulin} interpret their results as a demonstration
that the hydrophilic part of the molecules does not play an important
role in the water Hb dynamics.
Note, however, that the strong slowing down of the reorientation in the
hydrophobic hydration shells is not reproduced by MD 
simulations:\cite{Laage09,MYJPCB} for instance, in a classical MD study
(using SPC/E water) the reorientation of OH groups tangential to
hydrophobic methyl groups was calculated to slow down only by a 
factor of about 1.4 at low solute concentrations.\cite{Laage09}
On the basis of these observations and controversial interpretations,
it appears to be extremely useful to perform accurate first principles 
simulations
to study a system where hydrophilic groups are absent in such a way 
to investigate the effect of a hydrophobic group only.
In fact, MD simulations based on empirical force fields, 
specifically designed to reproduce selected experimental data
or a limited set of quantum chemically computed structures,
do not provide a description completely 
independent from reference data and the reliability of the results,
at conditions significantly different from those where the potential
was designed for, may be questionable.
These limitations can be overcome by Car-Parrinello
MD simulations, where the interactions are calculated from first principles,
and that have been already used in studies of liquid
water,\cite{Sprik96,PRLJCP,Schwegler,VandeVondele,Sit} aqueous 
solvation,\cite{MYJPCB,Marx,Erp,Erp03,Boero,Ikeda} and nucleic acid models
containing both hydrophilic and hydrophobic groups.\cite{Boero2,Boero3}
With respect to force-field based MD, first principles MD has 
important advantages:
it naturally incorporates polarization, it accounts for the
intramolecular motion, and gives detailed 
information on the electronic properties such as the charge distribution.

We have recently studied\cite{MYJPCB} the interaction of water with methanol 
(the simplest alcohol),
which can be considered as a prototype of a solute with both
hydrophobic (methyl) and hydrophilic (hydroxyl) groups: we found
that the presence of a methanol molecule slightly reduces
the mobility of many water molecules, rather than immobilizing
just the few ones which are closest to the methyl group, in line with
the interpretation of Qvist and Halle\cite{Qvist} and with
classical MD simulation results.\cite{Laage09,Ferrario}
Here we study, from first principles,
the interaction of water with methane 
which can be considered as a prototype of an hydrophobic solute. 
A previous first principles study of a water-methane solution
exists, which was however limited to the investigation of electronic
properties and polarization effects.\cite{Mateus} 

\section{Method}
Density Functional Theory (DFT)-based
Car-Parrinello (CP) MD simulations\cite{CPMD} have been performed
at constant volume, starting with the experimental density of water at 
room temperature.
The initial system consists of 64 equilibrated water 
molecules in a supercell
with body-centered-cubic symmetry\cite{bcc} and periodic boundary conditions.
Subsequently, two adjacent water molecules have been replaced by
a methane molecule, the resulting solution system being characterized 
by a density of 0.98 g/cm$^3$, in line with the typical density 
considered in previous simulations.\cite{Laaksonen,Mateus,Bridgeman} 
Then the novel system has been reequilibrated for
2 ps to allow water molecules to reorganize and form the solvation
shell around the solute.
Then statistics have been collected for about 20 ps, while the temperature
of the system has been controlled by a Nos\'e-Hoover
chain thermostat\cite{Nose} on the ionic degrees of freedom.  
For comparison with the properties of pure water, a parallel 
simulation with a system of 64 water molecules
at the same conditions, has been performed. 
In line with the approach followed in recent simulations,\cite{Sit,Sharma} 
the chosen target simulation temperatures were higher than the room temperature
of actual experiments, since standard
DFT functionals (such as the PBE adopted here) are
known to be affected by a slight overbinding leading to a sort of
``glassy behavior'' of the dynamics around room 
temperature,\cite{VandeVondele,Paesani11}
which is probably partially related to the neglect of 
proton quantum effects\cite{Schwegler} and to the well-known
deficiencies of standard DFT approaches in describing van der Waals 
interactions;\cite{Yoo11} 
note that the lack of an accurate description of the van der Waals 
interactions in the solvation of small hydrophobic molecules is
however expected to have little influence on the surrounding
water structure.\cite{REVIEW}  
According to extensive 
tests\cite{Sit,Paesani11,Yoo}
on liquid water simulations with the PBE functional, a 400 K
temperature guarantees a liquid-like dynamical behavior of the pure-water
system. 
In order to both directly verify this conclusion and also test its 
validity for aqueous solution systems, we have performed first principles
CP simulations by selecting target average temperatures of 330, 400, and 460 K
for pure water, and 400 and 460 K for the water-methane solution. 
H nuclei were treated as classical particles with the mass of 
deuterium which allows us to use larger time steps,\cite{Laasonen1} 
and the effective mass for the 
fictitious dynamics of the electrons was 700 a.u. with a MD 
time step of 0.121 fs; such an electronic fictitious mass
is within the range 
suggested ($\leq 370$ and $\leq 760$ a.u. for H$_2$O and D$_2$O,
respectively) to get an accurate description of the electronic
ground state and reliable structural and dynamical properties.\cite{Schwegler} 
The number (62) of solvent water molecules is expected
to be sufficient, since both experiments\cite{oldexpt} and previous 
simulations\cite{SIMULATIONS,Erp,Bakulin,Stirnemann,newsim}
indicate that solute-induced perturbations are short-ranged.
The computations were performed at the $\Gamma$-point only of the 
Brillouin zone, using norm-conserving Troullier-Martins\cite{Troullier} 
pseudopotentials with a 
plane-wave cutoff of 80 Ry and the gradient-corrected 
PBE density functional,\cite{PBE} which 
gives\cite{VandeVondele,Sit,Schwegler,Sharma} 
a good description of H bonding in liquid water.

\section{Results}

\subsection{Structural properties}
Concerning the structural properties, in Fig. 1 we show
our computed O-O pair correlation function, $g_{\rm OO}(r)$, of
pure water at the different temperatures considered in the
present simulations, while in Fig. 2 we report the $g_{\rm OO}(r)$
and $g_{\rm CO}(r)$ functions of the water-methane solution, at 400 and 460 K,
compared to the corresponding functions obtained for the water-methanol
solution at 400 K.\cite{MYJPCB} 
As expected, the amplitude of the fluctuations of all the pair correlation 
functions decreases by increasing the temperature; note that 
our $g_{\rm OO}(r)$ functions computed for pure water (Fig. 1) 
at 400 and 460 K are closer to
the experimental\cite{Soper} $g_{\rm OO}(r)$, obtained at room temperature, than
our $g_{\rm OO}(r)$ at 330 K, thus confirming that higher simulation
temperatures somehow mimic the missing van der Waals and nuclear effects. 
Moreover, the $g_{\rm OO}(r)$ functions of the solutions 
are more structured (Fig. 2) than that relative to pure water 
at the same temperature, in line with the conclusions of previous 
studies;\cite{Laaksonen,Bridgeman,Ferrario,Okazaki,Madura} 
a similar behavior can be also observed for the O-H and H-H 
pair correlation functions (not shown). 
The basic structural data (positions and height of the first maximum
and first minimum of the pair correlation functions, and coordination
number) are listed in Tables I and II, relative to the $g_{\rm OO}(r)$
and $g_{\rm CO}(r)$ function, respectively.
Similarly to the case of aqueous solvation of methanol,\cite{MYJPCB}
the distribution of water O atoms around the
C atom of methane is characterized by a broad first peak,
located between 3 and 5 \AA\ (see Fig. 2), 
indicating the existence of a solvation shell of about 20 water molecules,
in agreement with classical MD simulations.\cite{Laaksonen,Chau,Mateus,Dec}
The structure of the methane-water solution in the first hydration shell
of methane is determined by a competition between weak methane-water 
and significantly stronger water-water interactions.
\cite{Mateus}
A water molecule is defined\cite{Bridgeman} to be in the hydration shell 
if it is within the first minimum of the methane molecule (around 5.4 \AA);
bulk molecules are instead those outside the shell.
Using the ratio of heights of the first maximum and the following minimum 
of the $g_{\rm OO}(r)$ pair correlation function as
a measure of the degree of structure,\cite{Bridgeman} looking at Table I,
one can see that it obviously decreases for pure water by increasing
the simulation temperature; moreover, with respect to pure water,
at 400 K, it is a 2.2 factor larger for the
water-methane solution and a 1.3 factor larger for 
the water-methanol system, and, at 460 K, it is a 1.2 factor 
larger for the water-methane solution.  
As suggested elsewhere,\cite{Bridgeman} the enhanced structure of the
O-O pair correlation functions reflects the fact that water is excluded from 
the region occupied by the solute molecule and does not indicate an 
increase in water-water structure. 

The fraction of molecules participating in a given
number of Hbs has been also evaluated.\cite{HbCRITERION}
Clearly we focus attention on the water molecules located within
the solvation shell, this being a meaningful identification procedure
since the net displacement of both a typical solvent molecule and of the
solute during the simulation is relatively small.
In line with previous findings\cite{SIMULATIONS} 
the water molecules near the apolar methane molecule turn out to
have essentially the same average number of Hbs as do the others
(see Table V);
since the average distances and angles relative to these
Hbs are very similar to those in the bulk,
one infers that even the Hb strength is not significantly different.
Looking at Table I one can see that the nearest neighbor O-O distance
is essentially the same in pure water and in the solutions, if comparison
is made at the same simulation temperature. This confirms the 
results of neutron scattering studies which suggest that 
hydrophobic solutes have surprising little effect on the O-O 
distances,\cite{Bezzabotnov,Dixit} so that it appears that water molecules 
can change their Hb dynamics and partially lose
their reorientation ability without strongly disrupting the
local structure.\cite{Bakulin,Lazaridis}
The small changes in the structure of the solvating water shell 
around methane and methanol indicate that the Hb water network is
very flexible and can easily accommodate small apolar
solute groups without significantly changing its structure, in agreement
with previous studies.\cite{Laaksonen,Erp,Erp03}   
The low aqueous solubility of methane makes experimental studies of this
system difficult: estimates of the OC coordination number, 
$n_{OC}=$16 and 19 have been calculated from the
methane carbon-water oxygen pair correlation function determined
from neutron diffraction. Results from computer 
simulations\cite{Laaksonen,Lazaridis,Dec,Bridgeman,Chau} indicate that
$n_{OC}$ is in the range 16-22.
The chemical shifts in NMR spectra of aqueous methane can be used to
measure the number of water molecules in the first hydration sphere of 
methane because the chemical shift is a measure of the size of the
hydrophobic cage around methane: in such a way it was found that 
$n_{OC}=$20,\cite{Dec} in good agreement with our findings.

The orientations of the water molecules relative to the methane molecule
can be analyzed by computing the angle $\alpha$ subtended by the vector
joining the C atom of methane to the O of a water molecule with the 
vector between the O and a H atom of the water molecule.
For the water-methane solution, both at 400 and 460 K, 
we find $\alpha \simeq 91$ degrees, for the water molecules within the
solvation shell of methane, which corresponds to an essential
tangential orientation of water molecules with respect to the solute's 
surface. This type of arrangement agrees with previous classical MD 
estimates\cite{Chau,Tomlinson} and allows water molecules to straddle
the surface of the solute and maintain a nearly tetrahedral Hb 
coordination as found in bulk water.

\subsection{Dynamical properties}
More relevant to the comparison with recent experiments is the
analysis of the dynamical properties.
Our estimated diffusion coefficients (computed by mean-square displacement
and relative to the O atoms of 
the water molecules) in pure water and aqueous solutions are reported
in Table III. 
As can be seen, in pure water the diffusion coefficient, $D$, at the
simulation temperature of 400 K is close to the experimental value\cite{Lee}
of bulk liquid heavy water at room temperature, thus confirming that 
this simulation temperature is adequate to
reproduce effective room-temperature conditions (at the simulation 
temperature of 330 K $D$ is instead much smaller than the reference
value).
For the water-methanol solution at 400 K $D$ is only moderately
smaller than in pure water, in agreement with previous 
simulations\cite{SIMULATIONS} that predicted a slower translational 
diffusion of water in solution. 
However, for the water-methane system, at 400 K $D$ is much smaller
than the corresponding value of pure water at the same temperature
and is close to $D$ of pure water at 330 K, thus suggesting
that this strong water mobility reduction is probably not a genuine physical 
effect but instead a consequence of the artificial 
``glassy behavior'' of the dynamics discussed above.
At 460 K the $D$ value of water-methane exhibits instead only a
moderate reduction with respect to pure water at the same temperature,
similarly to the case of water-methanol at 400 K and in line with 
the common expectations. These observations suggest that the simulation
temperature of 460 K is more suitable for a proper comparison between 
the properties of the water-methane solution and those of pure water.

Among the available dynamical observables, the molecule
rotational correlation time represents a particularly useful
probe of the hydration shell because of its strong dependence 
on the local Hb configuration (in bulk water
molecular rotation occurs by a concerted mechanism where Hbs are
simultaneously broken and reformed\cite{Ramasesha}). In particular,
the rotational motion can be studied by looking at
the orientational correlation functions, given by

\begin{equation}
C_l(t)= < P_l[{\bf u}(t){\bf u}(0)] > \;,
\label{orient}
\end{equation}

where $P_l(x)$ is the Legendre polynomial of order $l$ and
${\bf u}$ is a unit vector along the molecular dipole moment
or the HOH angle bisector of the water molecules.
These orientational correlation functions can be related
to experimental measurements and different
techniques are sensitive to the first or second order dipole
correlation function:\cite{Tielrooij}
DR and Thz time domain spectroscopy
probe the decay of $C_1(t)$, while fs-IR (the technique used by 
Rezus and Bakker\cite{Bakker}), NMR, and optical Kerr-effect
spectroscopy probe the decay of $C_2(t)$. 
By assuming an exponential decay of the $C_1(t)$ and $C_2(t)$ functions
one can easily estimate, by a fitting and/or integration procedure
(see, for instance, ref. \onlinecite{Rosenfeld}), the corresponding
reorientation times, $\tau_1$ and $\tau_2$, which are reported in Table IV. 

In Fig. 3 we plot $C_2(t)$ relative to pure water and 
water-methane solution at 400 and 460 K (the corresponding figure, relative 
to water-methanol
was already reported and discussed in ref. \onlinecite{MYJPCB}).
As can be seen, all the functions exhibit an initial
rapid decay during the first half ps, which can be 
attributed\cite{SIMULATIONS,Tielrooij}
to overall molecular oscillation (libration) with a loss of phase
memory, but without significant net reorientation, keeping the
Hbs intact. 
By considering that the experimental anisotropy function
is given by $2/5\, C_2(t)$,\cite{Laage09}  
our computed $C_2(t)$ curve, at 460 K, resembles
the time-dependent anisotropy measured, at low
concentration, with pump-probe spectroscopy (see, for instance, Fig. 1 of 
ref. \onlinecite{Laage09} and Fig. 3 of ref. \onlinecite{Bakulin}).
In qualitative agreement with previous 
classical MD simulations,\cite{Laaksonen} the decay of $C_2(t)$, at 460 K, 
is found to be not much slower in solution than in 
pure water, without any noticeable plateau; instead, at 400 K, the
decay of our computed $C_2(t)$ turns out to be much slower in the
water-methane solution than in bulk water. 

The reorientation times (see Table IV), estimated from the
analysis of the $C_2(t)$ functions, relative to
all the water molecules, and, for the solutions, also to bulk waters, 
solvation-shell water molecules, and the 4 waters 
closest to methanol or methane
(roughly corresponding to integration of the $g_{\rm CO}(r)$ pair 
correlation function up to the position of the first maximum),
are only slightly longer than that in pure water
if comparison is done between water-methanol and pure water at 400 K
and between water-methane and pure water at 460 K.
In line with the behavior of the diffusion coefficient discussed above
and with the slow decay of the $C_2(t)$ function shown in Fig. 3,
the reorientation times of the water-methane solution at 400 K appear
to be instead much larger than in pure water, thus strengthining the conclusion
that the system is characterized by an artificial ``glassy behavior''
at that simulation temperature.
At the temperature of 460 K comparison with
pure water properties indicate a much more reasonable behavior,
thus indicating that the proper effective simulation temperature to be used
to reproduce ambient condition properties is not necessarily the same
for pure water and for solutions or confined water. 
As can be seen in Table IV, in the solutions 
there is no significant difference between
the retardation of the rotational motion of bulk waters and solvation waters;
only the 4 waters closest to methanol or methane appears to undergo a
little more pronounced slowing-down effect.
Our predicted moderate slowing down of water molecules around methane and
methanol is in good agreement with the results of classical MD 
simulations,\cite{Laaksonen,Laage09,Ferrario} although the absolute values of
our estimated reorientation times are significantly shorter than
those (2.8-4.5 ps)
obtained by some simulations\cite{Laage09,Ferrario} 
and in better agreement with the
data reported by Laaksonen and Stilbs\cite{Laaksonen} and with
experimental estimates (1.7-2.6 ps in pure water\cite{Paesani}),
thus confirming that reorientation rates are often
overestimated in classical MD simulations.\cite{SIMULATIONS}
Concerning the $\tau_1/\tau_2$ ratio, its value smaller than 3,
indicates that the reorientation motion of water molecules, both in bulk
and in solutions, cannot be entirely ascribed to a pure diffusive
mechanism, which would lead to a 3 value.\cite{Laage06}

\subsection{Hydrogen bond structure and dynamics}
In Table V we report the basic data about the Hb structure.
As can be seen, the average number of Hbs per water molecule, $N_{Hb}$,
slightly increases in solutions compared to the value of pure 
water at the same simulation temperature.
Moreover, as already pointed out above, the average number of the
four water molecules closest to the methane molecule does not much 
differ from (and in any case is not smaller than) that of the
other waters. 
The fraction of broken Hbs can be defined\cite{Chau} as 
$f=(4-N_{Hb})/2$, where $N_{Hb}$ is the average number of Hbs
per water molecule. We find that, at 400K, $f=0.26, 0.22, 0.11$
for pure water, water-methanol, and water-methane, respectively,
while, at 460K $f=0.35, 0.28$ for pure water and water-methane, respectively,
thus confirming again that small solutes like methanol and methane
not only do not disrupt the Hb structure of bulk water but they even
show a tendency to enhance it since the fraction of broken Hbs 
is slightly reduced in the aqueous solutions. 

In bulk liquid water the reorientation of water molecules
involves large angular jumps and proceeds through
a concerted mechanism in which a rotating water molecule breaks a Hb
with an overcoordinated first neighbor, while forming a new bond
with an undercoordinated molecule initially located on the inner side
of the second solvation shell (with typical energy barriers of the
order of 4 kcal/mol).\cite{Paesani11} 
Experiments and simulations for pure water have led to a detailed
picture of water motion at very short times (less than 100 fs),
followed by rotational jumps and Hb making and breaking at longer
times (of the order of 1-2 ps): the Hb exchange mechanism occurs by
means of large (about 60 degrees), very fast (about 200 fs) 
but relatively seldom (once in 2-3 ps) angular jumps corresponding to
concerted Hb switching events.\cite{Bakulin,Ramasesha,Rosenfeld,Laage06,Gaffney}
\cite{Skinner} 

In Table V we also list the average Hb lifetimes, estimated by
considering the persistence of all the Hb atom pairs during
the MD trajectory. 
As can be seen, the average Hb lifetimes of the four water molecules
closest to the methane are slightly longer (by about a
1.1 factor) than the value of all the waters.
Interestingly, the situation is different for aqueous solvation of
methanol, which is characterized by the presence of a hydrophilic group
too: in fact, the average Hb lifetime of the four water molecules 
closest to the CH$_3$ group of methanol is marginally shorter
that that of all the water molecules. 
The longer average Hb lifetimes at the interface relative to
the interior can be attributed to
the lack of cooperative effects in proximity to methane due to the
decreased density of water.\cite{Chakraborty} 
In fact, by considering the mechanism of Hb exchange,\cite{Rosenfeld}
in order to undergo Hb exchange, a water molecule must have 
a readily available Hb partner in the second hydration shell; at the
interface the number of nearest neighbor water molecules decreases; this
reduces the probability of there being a nearby water molecule that
may accept a newly formed Hb and thus lengthens the lifetime of
already formed Hbs.
Note that the rotational dynamics of water are inherently connected to the Hb 
dynamics, since for the rotational correlation to decay to
zero the water-water Hb network must lose all correlation with its 
former state. These reorientation times therefore are expected to be
roughly proportional to the Hb lifetimes,\cite{Rosenfeld}
since the Hb  exchange has been shown to be the main water
reorientation pathway.\cite{Laage06,Laage09,Sciortino}
This expectation is confirmed by our data if one compares the
$\tau_2$ values of Table IV to the $\tau_{Hb}$ estimates of
 Table V. 
Clearly, the presence of a hydrophobic group blocks the approach of some new
water Hb acceptors, thus retarding the jump rate between Hb 
acceptors for surrounding waters;\cite{Laage09} the
jump time increase with increasing solute concentration can be 
quantitatively described by a transition state excluded-volume 
model:\cite{Laage09} the slowdown factor with respect to the bulk is
$\rho_v = 1/(1-f)$, where $f$ is the fraction of transition state 
locations for the new acceptor which are forbidden by the solutes's
presence.   
\cite{Laage09,Stirnemann}

In the case of solvation of methane, a methane molecule excludes the
centres of water molecules from a spherical volume less than 
5 \AA\ across; this volume is small enough that its presence in water
requires no breaking of Hbs: water molecules can adopt orientations that
allow Hb pattern to go around the solute, and the extent to which
bonds are broken at any instant is similar to that in the pure liquid. 
\cite{REVIEW}
In the case of water-methane at 460 K, by estimating the excluded volume 
fraction and the consequent retardation factor,\cite{Laage09}
we obtain 0.26 and 1.35, in good agreement with the predictions (0.28
and 1.39, respectively) of Laage {\it et al.}\cite{Laage09}
Note that this retardation factor matches well with the slowing-down factor
(1.3) derived by the reorientation time analysis thus indicating
that the rotational slowing down is compatible with that one predicted
on the basis of the excluded volume fraction.
As discussed above,
compared to pure water, the water structure is slightly enhanced
in the presence of methane. Of course
the enhanced water structure in the first hydration shell implies longer
lifetimes for the Hbs and, evidently, hindered dynamics: the effect is
equivalent as lowering the temperature in the vicinity of the 
solute,\cite{Laaksonen} which could also partially explain the 
``glassy behavior'' of the water-methane solution at 400 K.

\subsection{Electronic (polarization) properties}
In contrast to simulations based on semiempirical 
potentials, first principles schemes yield detailed information also on
the electronic charge distributions, which  
in liquid and aqueous solvation systems 
can be conveniently elaborated using the Maximally  
Localized Wannier Function (MLWF) method.\cite{PRLJCP,Erp03,Boero,Wannier}
In condensed phases the selfconsistent internal electric field
polarizes the water and solute molecules, leading to a large increase of 
their molecular dipole moments.
Using the MLWF scheme,\cite{PRLJCP,Boero,Wannier} the 
dipole moment of a system can be simply calculated by assuming that
the electronic charge is concentrated in point charges located at the
positions of the Wannier function centers.

Our computed dipole moments are reported in Table VI.
In bulk liquid water the average value of the dipole moment 
per water molecule is enhanced by about 80 \% , with respect to
that of the isolated water monomer (1.88 D, experimental: 1.86 D),
this being a typical feature of strongly Hb molecules.
However, as can be seen, the presence of the solutes affects the distribution
of the water-molecule dipole moments: in fact, considering the 
associated statistical errors, the average dipole moment 
of water molecules in the solvation shells are significantly 
smaller than that of the bulk waters (the effect
is more pronounced for methanol, which is understandable since this
molecule is also characterized by the presence of a hydrophilic
termination). This suggests 
that the presence of the solute reduces the polarization
effects in the solvation-shell waters, which is compatible with the
fact that the rotational mobility is not much decreased:
in fact one expects that a smaller
dipole moment makes the reorientation of water molecules easier,
in such a way to partially offset the increased tendency to rigidity of
waters in the solvation shell (water rotation is very fast in the
limiting case of water dispersed in an apolar solvent\cite{Goodnough}).
Clearly such an effect cannot be reproduced by most of the
classical MD simulations where nonpolarizable models for the
water molecules (which are often assumed to be rigid and equal)
are adopted and could explain why the ratio between our estimated
reorientation times of waters in solution and in pure water is 
smaller than that (1.5-2.0) obtained by classical MD methods 
based on rigid force
field and instead closer to that (1.3) given by flexible-force field 
simulations.\cite{SIMULATIONS,Laage09} 

Recently, Mateus {\it et al.}\cite{Mateus} have investigated
the electronic properties
of a methane-water solution by a sequential quantum mechanical/MD approach
and found that, upon hydration, methane acquires an induced dipole moment
of $0.48 \pm 0.20$ D, due to polarisation effects and to weak methane-water
Hb interactions, a value comparable to our estimates in Table VI.
Mateus {\it et al.}\cite{Mateus} found no difference between the 
average monomeric dipole moment of bulk waters and that of waters in close
interaction (within the first hydration layer of) with methane;
instead we find a slight reduction. 
This difference is probably due the
fact that in the simulations of Mateus {\it et al.}  
the geometry of the water molecules was kept rigid 
while only that of methane was flexible, at variance with our approach.

\section{Conclusions}
In conclusion, our results, compatible with recent experiments,
shed light on the controversial issue related 
to the presence of ``immobilized'' water molecules around 
hydrophobic groups. Since there is no evidence that a few waters
rotate much more slowly than the others in solution, one concludes that 
the presence of the hydrophobic solute slightly reduces
the mobility of many water molecules, rather than immobilizing
just the few ones which are closest to the methane molecule, 
similarly to what
obtained by simulations of soluted methanol, and in line with
the interpretation of Qvist and Halle\cite{Qvist} and with 
classical MD simulation results.\cite{Laage09,Ferrario} 
Therefore our results are also relevant to validate the classical 
MD approach for modeling of aqueous solutions.
Moreover, 
the rotational slowing down is compatible with that one predicted 
on the basis of excluded volume fraction, which leads to a slower Hydrogen 
bond-exchange rate.
By generating maximally-localized Wannier functions, a detailed 
description of the polarization effects in both solute and solvent molecules
has been obtained, which better characterizes the solvation process.  
Finally, the analysis of
simulations performed at different temperatures suggests that 
the target temperature of the soluted system must be carefully chosen,
in order to avoid artificial slowing-down effects.

\acknowledgments
Allocation of computer resources from INFM
``Progetto Calcolo Parallelo'' is acknowledged.


\vfill
\eject

\begin{table}
\caption{First maximum and first minimum height (in parenthesis
position values in \AA) of the O-O pair correlation function,
their ratio (as a measure of the degree of structure\cite{Bridgeman}),
nearest neighbors O-O distance $d_{nn}$ (in \AA), and O-O 
coordination number, $n_{OO}$.}
\begin{center}
\begin{tabular}{|c|c|c|c|c|c|}
\hline
 system & 1st max.  & 1st min. & ratio & $d_{nn}$  & $n_{OO}$  \\ \tableline
\hline
 pure water (330 K)            & 3.32 (2.71) & 0.30 (3.25)& 11.1&2.81& 4.0 \\
 pure water (400 K)            & 2.54 (2.78) & 0.74 (3.32)&  3.4&2.80& 4.3 \\
 pure water (460 K)            & 2.29 (2.78) & 0.90 (3.39)&  2.5&2.87& 4.7 \\
 pure water (298 K, expt.$^a$) & 2.12---2.42 (2.76---2.86) & --- &  ---&2.80 & --- \\
\hline
 water-methanol (400 K)        & 2.73 (2.71) & 0.61 (3.32)&  4.5&2.81 & 4.2 \\
\hline
 water-methane (400 K)         & 3.33 (2.71) & 0.45 (3.32)&  7.4&2.82  & 4.2 \\
 water-methane (460 K)         & 2.39 (2.78) & 0.83 (3.32)&  2.9&2.86  & 4.3 \\
\hline
\end{tabular}
\tablenotetext[1]{ref.\onlinecite{Soper}.}
\label{goor}
\end{center}
\end{table}

\begin{table}
\caption{First maximum and first minimum height (in parenthesis
position values in \AA) of the C-O pair correlation function,
and C-O coordination number, $n_{OC}$.}
\begin{center}
\begin{tabular}{|c|c|c|c|}
\hline
 system & 1st max.  & 1st min.  & $n_{OC}$  \\ \tableline
\hline
 water-methanol (400 K)        & 1.75 (3.59) & 0.86 (5.76) &25.6 \\
\hline
 water-methane (400 K)         & 2.59 (3.79) & 0.52 (5.28) &19.8 \\
 water-methane (460 K)         & 1.91 (3.72) & 0.75 (5.55) &22.2 \\
 water-methane (298 K, ref.$^a$) & 1.84 (3.65) & 0.82 (5.35) &19.7 \\
 water-methane (320 K, ref.$^b$) & 1.43 (3.73) & 0.75 (5.4) &17.3 \\
\hline
\end{tabular}
\tablenotetext[1]{ref.\onlinecite{Mateus}.}
\tablenotetext[2]{ref.\onlinecite{Chau}.}
\label{gcor}
\end{center}
\end{table}

\begin{table}
\caption{Diffusion coefficient, $D$ in $10^{-5}$ cm$^2$/s,
(in parenthesis, in the case of the solutions, the reduction
factor with respect to $D$ of 
pure water at the same simulation temperature).}
\begin{center}
\begin{tabular}{|c|c|}
\hline
 system      & $D$  \\ \tableline
\hline
 pure water (330 K)            & 0.40 \\
 pure water (400 K)            & 2.60 \\
 pure water (460 K)            & 5.17 \\
 pure water (309 K, expt.$^a$) & 1.9 \\
\hline
 water-methanol (400 K)        & 1.66 (1.57)  \\
\hline
 water-methane (400 K)         & 0.34 (7.65) \\
 water-methane (460 K)         & 4.29 (1.21) \\
\hline
\end{tabular}
\tablenotetext[1]{ref.\onlinecite{Lee}.}
\label{diffusion}
\end{center}
\end{table}

\begin{table}
\caption{Reorientation times, $\tau_1$ and $\tau_2$ (in parenthesis
divided by the corresponding quantities of pure 
water at the same temperature), and $\tau_1/\tau_2$ ratio,
of all the water molecules, and, for the solutions, of bulk waters, BW, 
solvation-shell water molecules, SW, and the 4 waters closest 
to methanol or methane, 4W.}
\begin{center}
\begin{tabular}{|c|c|c|c|}
\hline
 system & $\tau_1$ (ps) & $\tau_2$ (ps) & $\tau_1/\tau_2$ \\ \tableline
\hline
 pure water (330 K)            & 15.8 &  5.9 & 2.7 \\
 pure water (400 K)            &  3.4 &  1.4 & 2.5 \\
 pure water (460 K)            &  1.5 &  0.6 & 2.6 \\
 pure water (300 K, classical MD$^a$) & 4.3 & 2.0 & 2.1\\
 pure water (283 K, classical MD$^b$) & 3.3 & 1.1 & 2.9\\
 pure water (300 K, expt.$^a$) & 2-7.5 &  1.7-2.6 & ---\\
\hline
 water-methanol (400 K)        &  3.8 (1.1) & 1.8 (1.3) & 2.1  \\
 water-methanol (400 K) BW     &  4.1 (1.2) & 1.9 (1.3) & 2.2  \\
 water-methanol (400 K) SW     &  4.3 (1.3) & 2.0 (1.4) & 2.2  \\ 
 water-methanol (400 K) 4W     &  5.2 (1.5) & 2.2 (1.6) & 2.4  \\
\hline
 water-methane (400 K)         & 10.6 (3.1) & 4.9 (3.6) & 2.1  \\
 water-methane (400 K)  BW     & 14.0 (4.1) & 5.8 (4.2) & 2.4  \\
 water-methane (400 K)  SW     &  9.1 (2.6) & 4.4 (3.2) & 2.1  \\
 water-methane (400 K)  4W     & 22.6 (6.6) & 8.6 (6.2) & 2.6  \\
\hline
 water-methane (460 K)         &  1.9 (1.3) & 0.7 (1.3) & 2.6  \\
 water-methane (460 K)  BW     &  1.9 (1.3) & 0.8 (1.4) & 2.4  \\
 water-methane (460 K)  SW     &  2.3 (1.5) & 0.7 (1.3) & 3.3  \\
 water-methane (460 K)  4W     &  2.4 (1.6) & 0.9 (1.6) & 2.7  \\
 water-methane (298 K, classical MD$^c$) & --- & 2.8-3.4 (1.3-1.5) & ---\\
\hline
\end{tabular}
\tablenotetext[1]{ref.\onlinecite{Laage06}.}
\tablenotetext[2]{ref.\onlinecite{Laaksonen}.}
\tablenotetext[3]{ref.\onlinecite{Laage09}.}
\label{tau}
\end{center}
\end{table}

\begin{table}
\caption{Average number of Hbs per water molecule, $N_{Hb}$,
average lifetime of water-water Hbs of all the water molecules,
$\tau_{Hb}$, and, for the solutions, of the 4 waters closest 
to methanol or methane, 4W (in parenthesis
divided by the corresponding quantities of pure 
water at the same temperature).}
\begin{center}
\begin{tabular}{|c|c|c|}
\hline
 system & $N_{Hb}$ & $\tau_{Hb}$ (ps)   \\ \tableline
\hline
 pure water (330 K)            & 3.79 &  2.37  \\
 pure water (400 K)            & 3.51 &  0.76  \\
 pure water (460 K)            & 3.33 &  0.47  \\
 pure water (300 K, classical MD$^a$) & 3.34-3.78 & 0.68-1.03\\
\hline
 water-methanol (400 K)        & 3.56 (1.02) & 1.01 (1.33)   \\
 water-methanol (400 K) 4W     & 3.56 & 0.99 (1.30)   \\
\hline
 water-methane (400 K)         & 3.71 (1.09) & 1.70 (2.24)   \\
 water-methane (400 K)  4W     & 3.80 & 1.86 (2.45)   \\
\hline
 water-methane (460 K)         & 3.38 (1.04) & 0.62 (1.32)   \\
 water-methane (460 K)  4W     & 3.40 & 0.69 (1.47)   \\
 water-methane (320 K, classical MD$^b$) & 3.52 & --- \\
\hline
\end{tabular}
\tablenotetext[1]{ref.\onlinecite{Bergman}.}
\tablenotetext[2]{ref.\onlinecite{Chau}.}
\label{lifetime}
\end{center}
\end{table}

\begin{table}
\caption{Average dipole moment, $\mu$, of the bulk water molecules, BW,
and, for the solutions, of the solvation shell waters, SW,
and of  methanol or methane.} 
\begin{center}
\begin{tabular}{|c|c|}
\hline
system & $\mu$(D)  \\ \tableline
\hline
 pure water (400 K)            & $3.49 \pm 0.16$ \\
 pure water (460 K)            & $2.94 \pm 0.01$ \\
\hline
 water-methanol (400 K) BW     & $3.56 \pm 0.17$ \\
 water-methanol (400 K) SW     & $3.05 \pm 0.02$ \\
 water-methanol (400 K) methanol&$3.05 \pm 0.10$ \\
\hline
 water-methane (400 K)  BW     & $3.16 \pm 0.02$ \\
 water-methane (400 K)  SW     & $3.05 \pm 0.02$ \\
 water-methane (400 K) methane &$0.26 \pm 0.03$ \\
\hline
 water-methane (460 K)  BW     & $2.99 \pm 0.02$ \\
 water-methane (460 K)  SW     & $2.94 \pm 0.02$ \\
 water-methane (460 K) methane &$0.34 \pm 0.04$ \\
\hline
\end{tabular}
\label{orientation}
\end{center}
\end{table}

\vfill
\eject

\pagestyle{empty}
\begin{figure}
{\vskip 1.3cm}
\centerline{
\includegraphics[width=17cm]{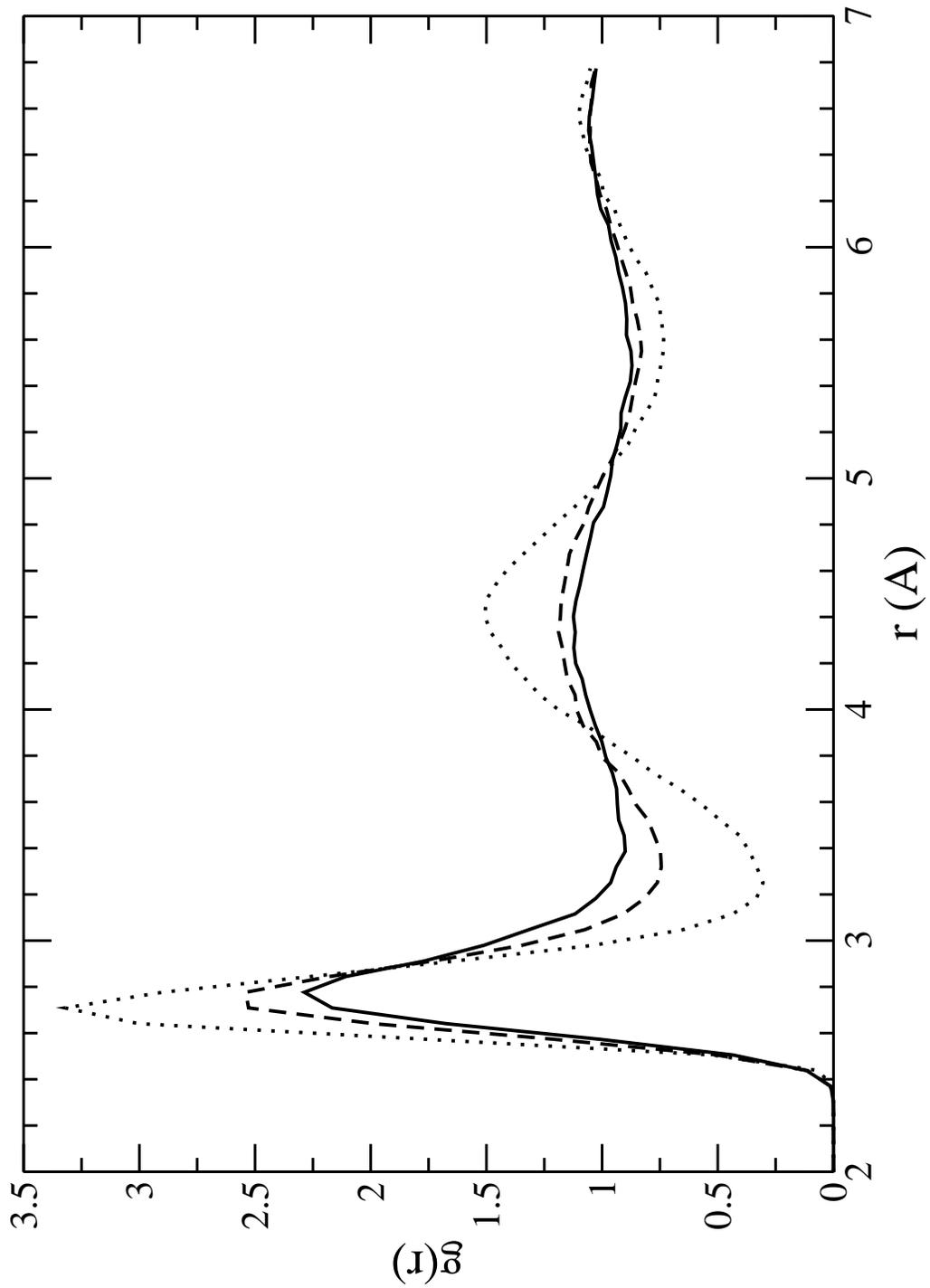}
}
\caption{O-O pair-correlation functions of pure water at different 
simulation temperatures: 330 (dotted line), 400 (dashed line), and 460 K
(solid line).} 
\label{fig1}
\huge
\end{figure}

\pagestyle{empty}
\begin{figure}
{\vskip 1.3cm}
\centerline{
\includegraphics[width=17cm]{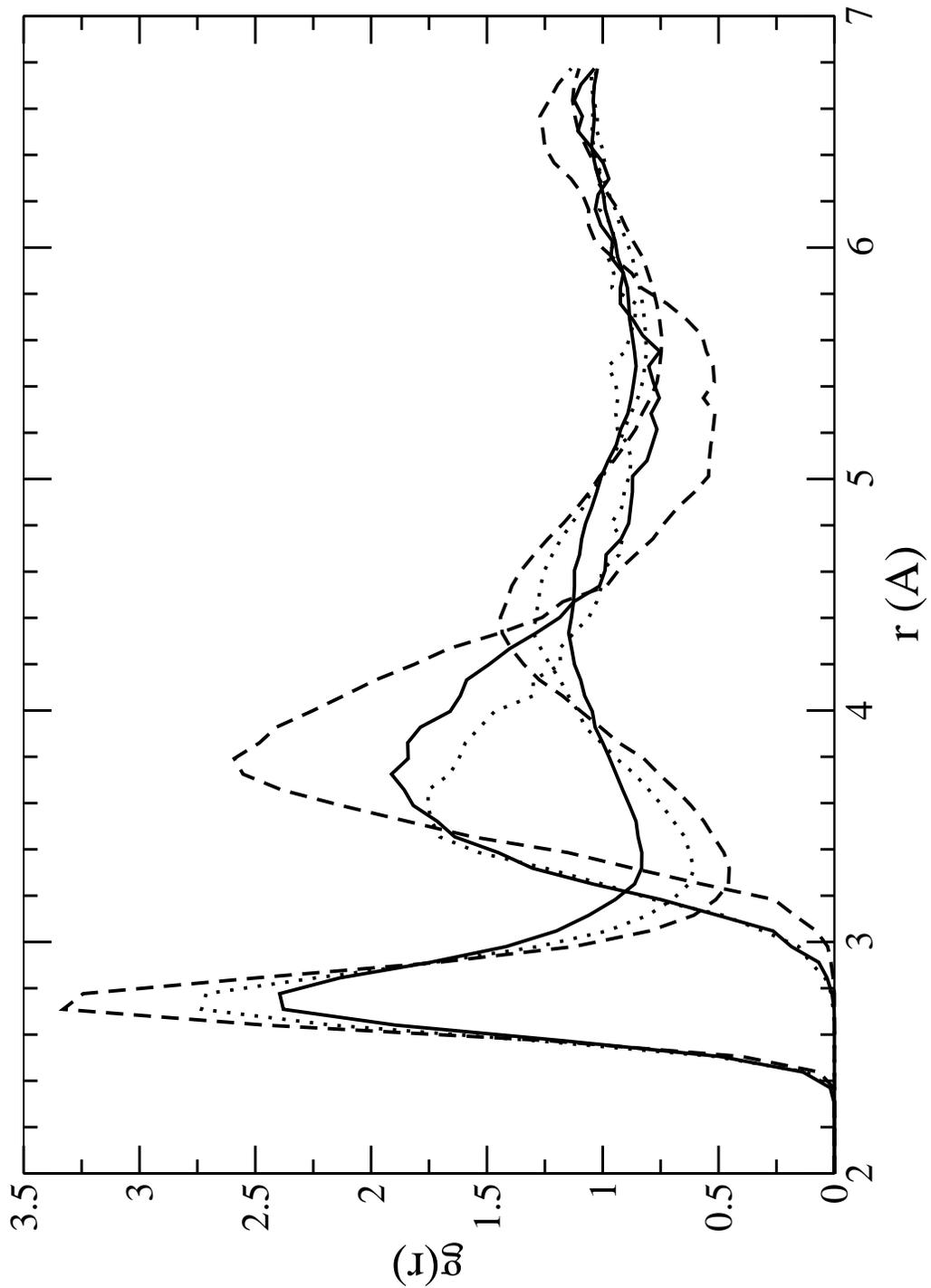}
}
\caption{O-O and C-O pair-correlation functions, with the main peak located around 2.8 and 3.5 \AA, respectively, of water-methane solution
at 400 (dashed line) and 460 K (solid line), and of water-methanol at 400 K
(dotted line).}
\label{fig2}
\huge
\end{figure}

\vfill
\eject

\pagestyle{empty}
\begin{figure}
{\vskip 1.3cm}
\centerline{
\includegraphics[width=17cm]{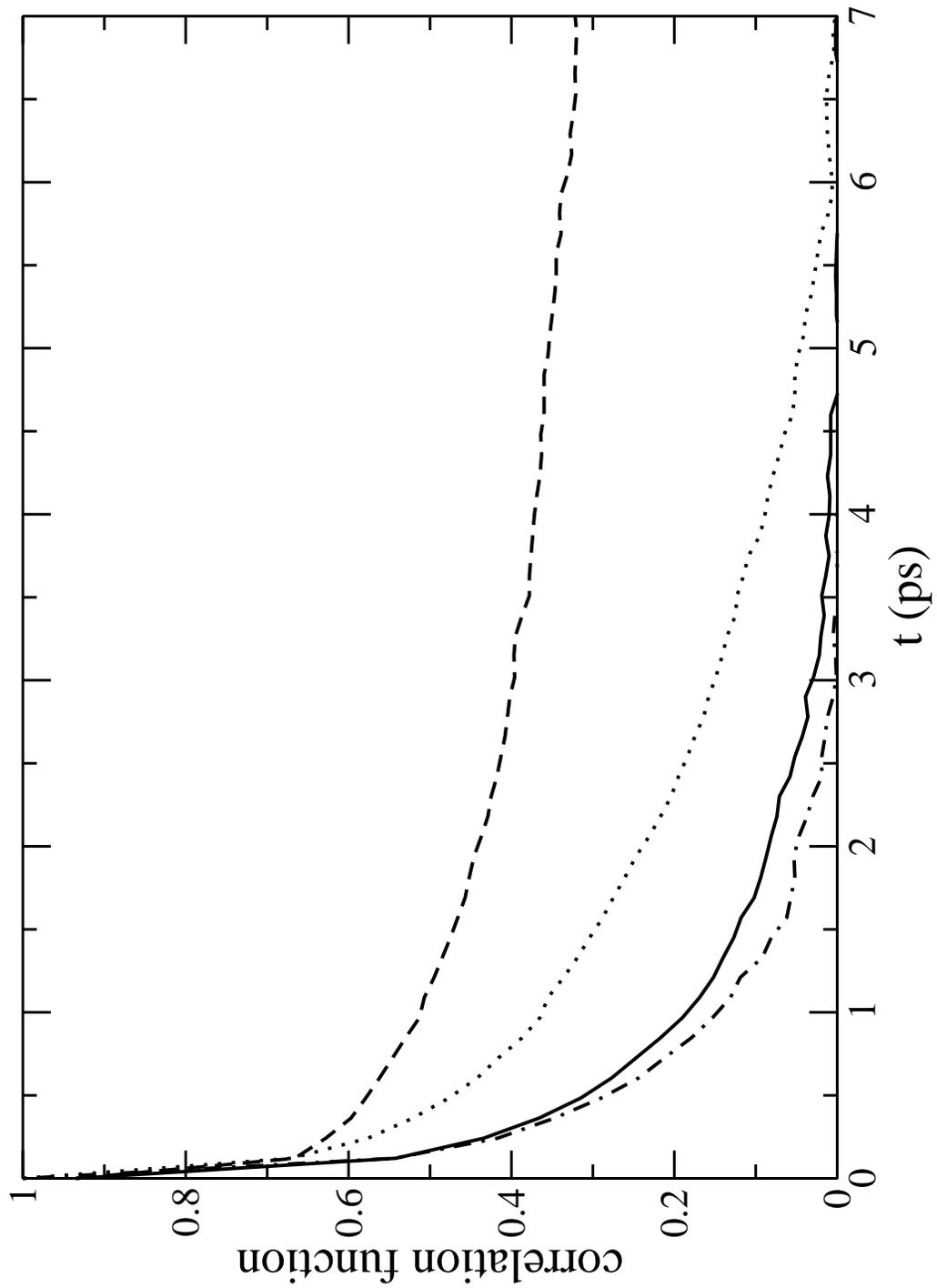}
}
\caption{$C_2(t)$ function in the water-methane solution at 400 (dashed
line) and 460 K (solid line), and in pure water at 400 (dotted
line) and 460 K (dot-dashed line).} 
\label{fig3}
\huge
\end{figure}

\vfill
\eject


\begin{references}
\bibitem{bcc} Such a ``truncated octahedron'' unit cell is often used in
              simulations of liquids since it more closely approaches 
              a sphere and is therefore more isotropic.
\bibitem{HbCRITERION} Here Hbs are identified
                      according to the following geometrical criterion: 
                      two molecules are
                      H bonded when the distance between the O atom
                      of one molecule and a H atom of the other is in
                      the range 1.5-2.4 \AA\ and the
                      O-H$\cdot \cdot \cdot$O angle
                      is in the range 120$^\circ$-180$^\circ$;
                      the results are not qualitatively changed by adopting 
                      other reasonable Hb criteria.
\end{references}
\end{document}